\documentclass[twocolumn,floatfix]{revtex4}
\usepackage{amsmath}
\usepackage{amssymb}
\usepackage{bm}
\usepackage{epsfig}
\usepackage{graphicx}
\usepackage{natbib}
\newcommand{\nix}[1]{}
\usepackage{color}
\begin{document}

\title{Spin coherence of a two-dimensional electron gas induced by
resonant excitation of trions and excitons in CdTe/(Cd,Mg)Te
quantum wells}
\author{E. A. Zhukov$^{1,2}$, D. R. Yakovlev$^{1,3}$, and M. Bayer$^1$}
\affiliation{$^1$Experimentelle Physik 2, Universit\"at Dortmund,
D-44221 Dortmund, Germany\\
$^2$Faculty of Physics, M.V. Lomonosov Moscow State University,
119992 Moscow, Russia\\
$^3$A.F. Ioffe Physico-Technical Institute, Russian Academy of
Sciences, 194021 St. Petersburg, Russia}

\author{M. M. Glazov and E. L. Ivchenko}
\affiliation{A.F. Ioffe Physico-Technical Institute,  Russian
Academy of Sciences, 194021 St. Petersburg, Russia}

\author{G. Karczewski, T. Wojtowicz, and J. Kossut}
\affiliation{Institute of Physics, Polish Academy of Sciences,
PL-02668 Warsaw, Poland}

\date{\today}

\begin{abstract}
The mechanisms for generation of long-lived spin coherence in a
two-dimensional electron gas (2DEG) have been studied
experimentally by means of a picosecond pump-probe Kerr rotation
technique. CdTe/(Cd,Mg)Te quantum wells with a diluted 2DEG were
investigated. The strong Coulomb interaction between electrons and
holes, which results in large binding energies of neutral excitons
and negatively charged excitons (trions), allows one to address
selectively the exciton or trion states by resonant optical
excitation. Different scenarios of spin coherence generation were
analyzed theoretically, among them the direct trion photocreation,
the formation of trions from photogenerated excitons and the
electron-exciton exchange scattering. Good agreement between
experiment and theory is found.
\end{abstract}

\maketitle
\section{Introduction}

The spin coherence of electronic states is one of the key features
involved in numerous concepts for spintronics devices (see, e.g.,
Refs.~[\onlinecite{Aws02}, \onlinecite{Zut04}]). It has been
studied in semiconductor structures of different dimensionality,
including bulk-like thin films, quantum wells (QWs) and quantum
dots. Accordingly, the spin coherence time of an electron has been
found to vary over a wide range from a few picoseconds up to a few
microseconds.

An ensemble of electron spins subject to a magnetic field is
commonly characterized by three relaxation times \cite{Abr}. The
longitudinal spin relaxation time $T_1$, which is related to the
relaxation of the spin component parallel to the field, and the
transverse relaxation times $T_2$ and $T_2^*$. The $T_2$ time
describes spin decoherence (i.e. relaxation of the spin components
transverse to the field) of a \emph{single} electron, while the
$T_2^*$ time describes the dephasing of the whole spin ensemble.

The coherence time $T_2$ of a single spin is often few orders of
magnitude longer than the dephasing time $T_2^*$ of a spin
ensemble. For example, in (In,Ga)As/GaAs quantum dots these times
are 3 $\mu$s and 0.4 ns, respectively, at $B$ = 6 T
\cite{Gre06,Gre06b}. A $T_2^*$ of 300 ns has been measured in bulk
GaAs by means of the Hanle effect on optically oriented electrons
at liquid helium temperature \cite{Dzh01}. For QWs the longest
spin dephasing times reported so far are 10 ns for GaAs/(Al,Ga)As
\cite{Dzh02} and 30 ns for CdTe/(Cd,Mg)Te \cite{Zhu06a,Hof06}.
They have been measured for structures with a very diluted
two-dimensional electron gas (2DEG). Electron spin coherence in
CdTe/(Cd,Mg)Te QWs attracts recently an increasing interest
\cite{Gil99,Zhu06a,Ast06,Hof06,Bra06,Che07,Tri07,Yak07}.

Figure 1 illustrates three typical situations realized in
experiments on coherent spin dynamics under resonant optical
excitation of the QWs. These cases are different in the density of
resident 2D electrons, $n_e$. In the undoped samples [panel (a),
$n_e=0$] spin oriented excitons are photogenerated. Usually hole
spin dephasing occurs within several picoseconds and the coherent
response is mainly determined by the spin dynamics of the electron
in the exciton. This dynamics cannot be studied for time scales
exceeding the exciton lifetime as the exciton recombination
depopulates the photoexcited states. For the dense 2DEG [panel
(c), $n_e a^2_B > 1$, where $a_B$ is the exciton Bohr radius],
exciton formation is suppressed because of state-filling and
screening effects. After photogeneration a hole looses its spin
and energy quickly and recombines with an electron from the Fermi
see. However, the spin oriented electron photogenerated at the
Fermi level has an infinite lifetime which allows one to study its
long-lived spin coherence and spin relaxation. In this case a
circularly polarized photon can increase the spin polarization of
the 2DEG by $S = \pm 1/2$. In case of a diluted 2DEG [panel (b),
$n_e a^2_B \ll 1$] the mechanism for generation of the electron
spin coherence is not so obvious. Indeed the ground state is a
singlet trion with an antiparallel orientation of electron spins.
Being resonantly excited, this state would not contribute to the
spin polarization because the hole undergoes fast decoherence and
the total spin of the two electrons is $S=0$.

\begin{figure}[htb]
\centering
\includegraphics[width=1.1\linewidth]{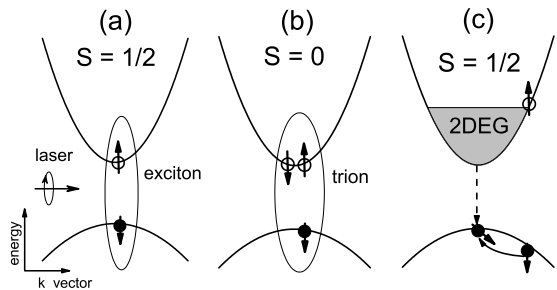}
\caption{Schematic presentation of generation of carrier spin
coherence by circular polarized laser pulses. The three cases
differ in the density of 2DEG in the QW: (a) empty QW, only
photogenerated carriers, which form excitons; (b) low density
2DEG, trions with a singlet ground state formed by a
photogenerated exciton and a background electron. Interaction of
the trion with the 2DEG is negligible; (c) dense 2DEG with a Fermi
energy exceeding the exciton binding energy. Excitons and trions
are suppressed by the 2DEG.}\label{fig:A1}
\end{figure}

However, generation of electron spin coherence has been observed
experimentally under resonant excitation of trions in QWs
\cite{Zhu06a,Ken06} and quantum dots \cite{Gre06}. There are two
equivalent approaches to explain the generation mechanism in this
case. The first one suggests that a coherent superposition of
electron and trion states is excited by a circular polarized
photon when the system is subject to an external magnetic field
\cite{Gre06,Ken06,Sha03}. The second one considers the 2D
electrons captured for trion formation: under circular polarized
excitation electrons with a specific spin orientation will be
extracted from the 2DEG and correspondingly spin polarization with
the opposite sign will be induced \cite{Zhu06a}.

In this paper we report on a detailed study of the coherent spin
dynamics of electrons and electron-hole complexes in
CdTe/Cd$_{0.78}$Mg$_{0.22}$Te QWs with a low density 2DEG,
performed by a pump-probe Kerr rotation (KR) technique. A
theoretical analysis of the various mechanisms of spin coherence
generation is carried out.

The paper is organized as follows. Sec. II is devoted to (i) the
theoretical analysis of the physical mechanisms leading to Kerr
and Faraday rotation of the probe pulse polarization, and (ii) the
theoretical description of various scenarios of electron spin
coherence generation. The experimental results are presented in
Sec. III, which also contains a comparison with the model
considerations of Sec. II. The paper is concluded by Sec. IV in
which we also comment on the specifics of \textit{p}-doped QWs.

\section{Theory: Model considerations}

Below we underline the two main aspects of the developed theory,
namely, the nature of Kerr rotation (KR) and Faraday rotation (FR)
signals and the models of spin dynamics of a 2DEG and
electron-hole complexes, both trions and excitons.

In short, the basic features of an experiment designed for
time-resolved measurements of the electron spin coherence can be
summarized as follows: the sample containing a 2DEG is excited
along the structure growth axis ($z$-axis) by an intense pump
pulse which induces resonant interband transitions. Then a much
weaker, linearly-polarized probe pulse with the frequency either
coinciding with or different from the pump frequency arrives at
the sample. The rotation of the polarization plane of the
reflected or transmitted probe pulse is analyzed as a function of
the delay between the pump and probe pulses. An external magnetic
field $B$ is applied in the QW plane, say, along the $x$-axis and
leads to precessions of the $z$ and $y$ electron spin components
with the Larmor frequency $\Omega \equiv \Omega_x = g_e \mu_B
B/\hbar$, where $g_e$ is the electron in-plane $g$-factor along
the $x$-direction and $\mu_B$ is the Bohr magneton. For 2D heavy
holes bound into excitons or trions, the in-plane $g$-factor is
very small and can be ignored \cite{Mar99}.

\subsection{Kerr and Faraday rotation in QWs}

In order to describe the KR in the pump-probe experiment under
normal-incidence resonant excitation we first consider the
amplitude reflection coefficient of an axially-symmetric single
QW, which in the vicinity of the exciton or trion resonance is
given by~\cite{ivchbook}
\begin{equation} \label{unpolarized}
r_{QW}(\omega) = \frac{{\rm i} \Gamma_{0}}{\omega_{0} - \omega -
{\rm i}(\Gamma_{0} + \Gamma)}\:,
\end{equation}
where $\omega$ is the incident light frequency and $\omega_{0}$,
$\Gamma_{0}$ and $\Gamma$ are the exciton (trion) resonance
frequency, the radiative and the nonradiative damping rates,
respectively. Taking into account also the cap layer as
constituent of the heterostructure, the total amplitude reflection
of the light incident on the structure from vacuum reads
\begin{equation}
\label{r} r = \frac{r_{01} + r_{QW} {\rm e}^{2 {\rm i} \phi}} {1 -
r_{10} r_{QW} {\rm e}^{2 {\rm i} \phi}}\:,
\end{equation}
Here $r_{01}= - r_{10} = (1 - n_b)/(1 + n_b)$ is the reflection
coefficient at the boundary between the cap layer and vacuum,
$n_b$ is the refractive index of the cap layer which for
simplicity is assumed to coincide with the background refractive
index of the well material, $\phi = k_b b$, $b$ is the cap-layer
thickness and $k_b$ is the light wave vector in the cap-layer. The
above equation is valid for a spin-unpolarized system in which
case the coefficient $r$ is insensitive to the light polarization.
For the spin-polarized resident electrons the reflection
coefficient for right ($+$) and left ($-$) circularly polarized
light has the form of Eq.~(\ref{unpolarized}) but $r_{QW}$ is
replaced by
\begin{equation} \label{reflection}
r_{QW, \pm}(\omega) = \frac{{\rm i} \Gamma_{0,\pm}}{\omega_{0,\pm}
- \omega - {\rm i}(\Gamma_{0,\pm} + \Gamma_{\pm})}
\end{equation}
with the parameters $\omega_{0}, \Gamma_{0}$ and $\Gamma$
dependent on the light helicity. These parameters are, in general,
determined by the concentration and spin polarization of the
carriers and their complexes as well as by the delay between the
pump and probe pulses. When describing the Faraday rotation effect
we have to analyze the amplitude transmission coefficient $t_{QW,
\pm}(\omega) = 1 + r_{QW, \pm}(\omega)$. Both the two signals
measured in reflection (Kerr rotation) or transmission (Faraday
rotation) geometry are proportional to the difference
\begin{equation} \label{signal}
\Sigma_{+} - \Sigma_{-} = \Sigma_0 {\rm Im}\{ r_+^* r_-\}\:,
\end{equation}
where $\Sigma_0$ and $\Sigma_{\pm}$ are the time-integrated
intensities of the incident and reflected/transmitted probe pulses
in time-resolved experiments (or the stationary intensities under
steady-state photoexcitation). Let us introduce the symmetrized,
$\bar{r}$, and antisymmetrized, $\tilde{r}$, combinations of
$r_{\pm}$ and assume $\tilde{r}$ to be small. Then in first
approximation one obtains
\begin{equation} \label{approxr}
{\rm Im}\{ r_+^* r_-\} = \frac{-2(1 - r_{10}^2)}{|1 - r_{10}
\bar{r} {\rm e}^{2 {\rm i} \phi}|^2}\ {\rm Im} \left[
\frac{\tilde{r} (r_{01} {\rm e}^{2 {\rm i} \phi} + \bar{r}^* )}{1
- r_{10} \bar{r} {\rm e}^{2 {\rm i} \phi}} \right]\:.
\end{equation}
If $\bar{r}$ is also small as compared with $r_{01}$ the
right-hand side of the above equation reduces to $-2r_{10}(1 -
r_{10}^2){\rm Im}\{ {\rm e}^{2 {\rm i} \phi} \tilde{r}  \}$.

The probe at the trion resonance frequency is described by
Eqs.~(\ref{r})--(\ref{signal}), where $\Gamma_{0, \pm}$ is the
oscillator strength for resonant trion excitation by
$\sigma_{\pm}$ circularly polarized light~\cite{Ast00}. For
heavy-hole optical transitions the values of $\Gamma_{0, \pm}$ are
proportional to the density of resident electrons with the spin
component $\pm 1/2$, respectively.

In the case of a dense 2DEG the Kerr or Faraday rotation angle is
proportional to~\cite{aronovivchenko}
\[
\left( \frac{e \hbar |p_{cv}|}{m_0 E_g^{QW}} \right)^2 {\rm
Re}\left\{ {\rm e}^{2 {\rm i} \phi}\int\limits_0^{\infty} d
\varepsilon \frac{ \mathcal D [ f_-(\varepsilon_e) -
f_+(\varepsilon_e)]}{E_g^{QW} + \varepsilon - \hbar \omega - {\rm
i} \hbar \Gamma_{eh}} \right\}\:.
\]
Here $p_{cv} = \langle S | \hat{p}_x | X \rangle$ is the interband
matrix element of the momentum operator, $E_g^{QW}$ is the band
gap renormalized by the quantum confinement of conduction
electrons and heavy holes, $\Gamma_{eh}$ is the damping rate for
an electron-hole pair, $m_0$ is the free electron mass,
$\varepsilon$ is the sum of the 2D electron and hole kinetic
energies, $\varepsilon_e = (\mu/m_e) \varepsilon$; $\mu = m_e
m_{hh}/ (m_e + m_{hh})$, $m_e$ and $m_{hh}$ is the reduced,
electron and heavy-hole effective mass, respectively,
$f_{\pm}(\varepsilon_e)$ is the energy distribution function for
electrons with spin $\pm 1/2$, and $\mathcal D$ is the reduced
density of states proportional to $\mu / \hbar^2$.

In the following we discuss, one after another, physical
mechanisms of the pump-probe signal Eq. (\ref{signal}) for three
cases of interest: (i) a diluted 2DEG subject to resonant circular
polarized photoexcitation in the singlet trion state, (ii) a
diluted 2DEG with resonant generation of excitons at low
temperatures favoring the binding of excitons and resident
electrons into trions, (iii) exciton generation in a dense 2DEG,
and (iv) a diluted 2DEG with photogeneration of carriers with
kinetic energy considerably exceeding the exciton binding energy.

\subsection{Resonant excitation of trions}\label{trion}

\begin{figure}[htbp]
\includegraphics[width=1\linewidth]{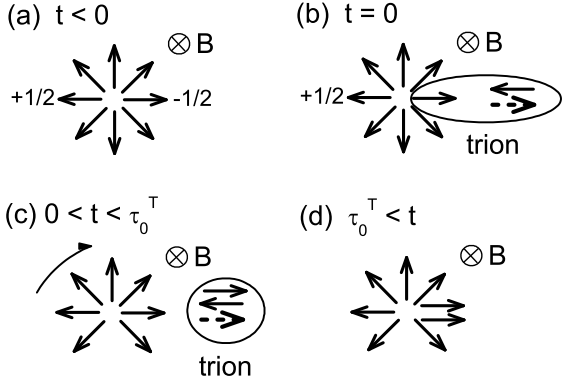}
  \caption {Scheme of generation of 2DEG spin coherence in external
magnetic fields via resonant photogeneration of trions. (a)
Initial state of a 2DEG which polarization in the plane
perpendicular to the magnetic field is zero. These spins are
precessing around $B$. (b) $\sigma^-$ polarized photon generates a
$(e, +1/2; hh, -3/2)$ electron-hole pair, which captures a $-1/2$
resident electron to form a trion. The 2DEG became polarized due
to uncompensated $+1/2$ electron spin left. (c) During trion
lifetime, $\tau^T_0$, the 2DEG polarization precesses around the
magnetic field. Trion state does not precess in magnetic field as
on the one hand its electronic configuration is singlet and on the
other hand in-plane hole g-factor is zero. (d) After trion
recombination the $-1/2$ electron is returned to the 2DEG (we
neglect here spin relaxation of the hole in the trion). Final
state of the 2DEG with induced polarization is
shown.}\label{fig:trion}
\end{figure}

We start the analysis from the case of KR by a QW with a diluted
2DEG and for resonant trion generation. In this case the radiative
homogeneous broadenings $\Gamma_{0,\pm}$ (coinciding with the
oscillator strength of the trion) in Eq.~\eqref{reflection} are
proportional to the concentrations of the resident electrons with
spin-down and spin-up, see Refs.~[\onlinecite{Ast00,Ast02}].
Qualitatively, the physical picture looks as follows. According to
the selection rules, the absorption of a circularly polarized
photon leads to the formation of an electron-hole pair with fixed
spin projections: $(e, -1/2; hh, +3/2)$ and $(e, +1/2; hh, -3/2)$
for right ($\sigma^+$) and left ($\sigma^-$) circularly polarized
photons, respectively. At weak and moderate magnetic fields the
ground state of negatively charged trions has a singlet electron
configuration with antiparallel orientation of electron spins.
Thus, for resonant excitation only resident electrons with
orientation opposite to the photogenerated electrons can
contribute to trion formation. This means that the 2DEG becomes
depleted of electrons with $z$-spin component $S_z = +1/2$ under
$\sigma^+$ pumping and of $S_z = -1/2$ electrons for $\sigma^-$
pumping. An external magnetic field applied in the plane of the
structure leads to precession of the spin polarization of resident
electrons and, therefore, to a modulation of $\Gamma_{0,\pm}$ and
oscillations of the Kerr signal. This process is shown
schematically in Fig.~\ref{fig:trion}.

Now we turn to an analytical description of this scenario. The
kinetic equations describing the spin dynamics of electrons and
trions after resonant, pulsed excitation of trions have the form
\begin{eqnarray} \label{1}
&&\frac{d S_z}{dt} = S_y \Omega - \frac{S_z}{\tau_s} +
\frac{S_T}{\tau^T_0 }\:, \\
&&\frac{d S_y}{dt} = - S_z \Omega - \frac{S_y}{\tau_s} \:, \nonumber \\
&&\frac{d S_T}{dt} =  - \frac{S_T}{\tau^T} \:. \nonumber
\end{eqnarray}
Here $S_T= (T_+ - T_-)/2$ is the effective trion spin density with
$T_{\pm}$ being the densities of negatively-charged trions with
the heavy-hole spin $\pm 3/2$, $S_y$ and $S_z$ are the
corresponding components of the electron-gas spin density,
$\tau^T$ is the lifetime of the trion spin including the trion
lifetime $\tau^T_0$ and the spin relaxation time $\tau^T_s$, i.e.,
$\tau^T = \tau_0^T \tau_s^T /(\tau_0^T + \tau_s^T)$, and $\tau_s$
is the electron spin relaxation time. It can be identified as the
ensemble transverse spin relaxation time $T_2^*$. Under normal
incidence of the $\sigma^+$ polarized pump the initial conditions
are $S_y(0) =0, S_T(0) = - S_z(0) = n_0^T/2$ with $n_0^T$ being
the initial density of photogenerated trions. Remember that the
magnetic field is directed along the $x$ axis, therefore the $x$
component of electron spin density is conserved.

Let us introduce the complex function $S^+ (t) = S_z(t) + {\rm i}
S_y(t)$. It satisfies the equation
\begin{equation} \label{s+}
\frac{d S^+}{dt} = - \left( \frac{1}{\tau_s} + {\rm i} \Omega
\right) S^+ + \frac{S_T}{\tau_0^T}
\end{equation}
with the initial condition $S^+(0) = - n_0^T/2$. The solution for
$S_T(t)$ is readily written as $ S_T(t) = \frac{1}{2}n_0^T {\rm
e}^{- t/\tau^T}$. Substituting $S_T(t)$ into Eq.~(\ref{s+}) one
finds
\begin{equation} \label{sta}
S^+(t) = - \frac{n^T_0}{2} [ (1 - \eta) {\rm e}^{ - ( \tau_s^{-1}
+ {\rm i} \Omega ) t } + \eta {\rm e}^{ - t/\tau^T}] \:,
\end{equation}
where $\eta = (\tau^T_0)^{-1}/[(\tau^T)^{-1} - \tau_s^{-1} - {\rm
i} \Omega]$. The real part of $S^+(t)$ is equal to $S_z(t)$ and
given by
\begin{equation} \label{sza}
S_z(t) = \frac{n_0^T}{2} [ |1 - \eta| \sin{(\Omega t - \Phi)} {\rm
e}^{ - t/ \tau_s }  - \eta' {\rm e}^{ - t/\tau^T } ]\:,
\end{equation}
where
\begin{equation}\label{phaseNEW}
\Phi = \arctan[{(1- \eta')/\eta''}],
\end{equation}
$\eta'$ and $\eta''$ are the real and imaginary parts of $\eta$.
In the particular case when electron and trion spin relaxation can
be ignored, Eq.~(\ref{sza}) is reduced to
\begin{equation} \label{sza'}
S_z(t) = \frac{n_0^T}{2}  [- \cos^2{\Phi}\ {\rm e}^{ - t/\tau^T }
+ \sin{\Phi}\ \sin{(\Omega t - \Phi)}]\:,
\end{equation}
where $\Phi = \arctan{(\Omega \tau^T)}$.

In s strong magnetic field such that $\omega = \Omega T^* \gg 1$
[with ${(T^*)}^{-1} ={(\tau^T)}^{-1} - \tau_s^{-1}$] the parameter
$\eta \to 0$ and $\sin{\Phi} \to 1$, $\cos{\Phi} \to 0$ yielding
$\Phi \to \pi/2$. The high field asymptotics reads
\begin{equation}\label{highfield}
\Phi \approx \frac{\pi}{2} -
\frac{1}{\Omega \tau_0^T}.
\end{equation}
For low magnetic fields $\Omega T^* \ll 1$  so that we have
\begin{equation}\label{smallfield}
\Phi \approx
\frac{\pi}{2} - \frac{\Omega (T^*)^2}{\tau_0^T - T^*}\:.
\end{equation}

The dependence of the phase $\Phi$ on magnetic field has a minimum
\begin{equation}\label{phase_min}
\Phi_{\rm min} = \arctan{\left(2\Omega_{\rm min}\tau_0^T\right)}.
\end{equation}
reached at
\begin{equation}\label{omega_min}
\Omega_{\rm min} = \frac{1}{T^*}\sqrt{1-\frac{T^*}{\tau_0^T}}\:.
\end{equation}

Note that Eqs. \eqref{omega_min} and \eqref{phase_min} are valid
provided that $\tau_s$ is magnetic field independent. In
principle, this might not be the case in the experiment, where the
main contribution to the ensemble dephasing time arises from the
spread of $g$-factor values, but for typical conditions $\tau_s
\gg \tau_0^T$ so that $T^* \approx \tau_0^T$ and is almost field
independent.

\begin{figure}[htb]
\centering
\includegraphics[width=1\linewidth]{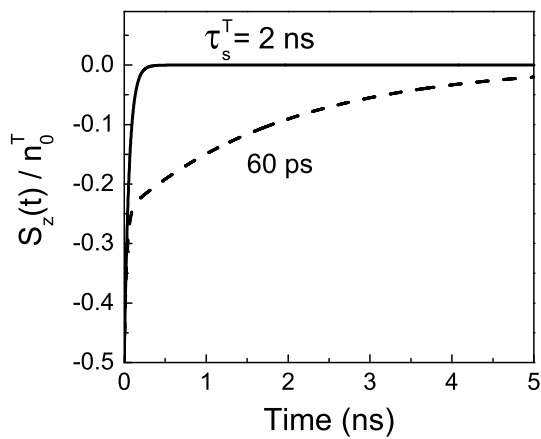}
\caption{Kerr signal calculated for pump and probe resonant with
the trion in absence of an external magnetic field, $B=0$. The
electron spin relaxation time $\tau_s = 2$~ns, the trion radiative
lifetime $\tau_0^T=60$~ps. Dashed curve corresponds to the trion
spin relaxation time $\tau_s^T=60$~ps, while solid curve
corresponds to an unrealistically high $\tau_s^T=2$~ns. The latter
situation corresponds to the compensation of the initial spin
polarization by the returned electrons, $\eta_0=1$.}
\label{fig:comp:theor}
\end{figure}

At zero magnetic field, the electron spin exhibits no precession,
$S_y \equiv 0, S^+(t) = S_z(t)$, and one has
\begin{equation} \label{st0}
S_z(t) = - \frac{n_0^T}{2}  [ \eta_0\ {\rm e}^{ - t/\tau^T} + (1 -
\eta_0)\ {\rm e}^{ - t/\tau_s } ] \:,
\end{equation}
where $\eta_0=\eta(B=0)$. This can be understood as follows. At
the moment right after photoexcitation into the trion resonance by
a pulsed, say, $\sigma^+$ polarized light, the system contains
$n_0^T$ singlet trions with completely polarized holes $+3/2$, and
$n_0^T$ electrons with uncompensated spin $-1/2$, because the same
number of electrons with spin $+1/2$ were extracted from the 2DEG
to form trions. This stage is illustrated by Fig.~\ref{fig:trion}
(b). In the absence of spin relaxation, trions decay by emitting
$\sigma^+$ photons and giving back lent electrons with spin
$+1/2$. As a result the initially generated electron spin
polarization is compensated by the returned electrons and tends to
zero as the trions vanish. The compensation takes place for the
limiting case $\tau^T_0 \ll \tau^T_s, \tau_s$ and also for the
special case of $\tau_s = \tau^T_s$, see solid curve in
Fig.~\ref{fig:comp:theor}. If the spin relaxation times $\tau_s$
and $\tau^T_s$ coincide, then $\eta_0=1$ and the second
contribution to $S_z(t)$ proportional to ${\rm e}^{ - t / \tau_s
}$ disappears. In this case the electron polarization $S_z(t)$
decays with the trion spin lifetime $\tau^T$.

Spin relaxation of any subsystem brings in an imbalance, namely,
the spins of the left over and the returned electrons cannot
completely compensate each other, see dashed curve in
Fig.~\ref{fig:comp:theor}. The two characteristic parts of the
dashed line correspond to the fast and slow negative contributions
to $S_z(t)$ in Eq.~(\ref{st0}) governed by $\tau^T = 30$ ps and
$\tau_s = 2$ ns, respectively.

In an in-plane external magnetic field the imbalance arises even
if spin relaxation is absent or $\tau_s = \tau^T_s$. Indeed, as
schematically illustrated in Fig.~\ref{fig:trion}(b), both the
heavy-holes (due to their zero in-plain heavy-hole $g$-factor) and
the singlet electron pairs bound in trions are not affected by the
magnetic field, whereas the spins of the resident electrons
precess. The trion recombines radiatively by emission of a
$\sigma^+$ photon and the returned electron is spin-up polarized
along the $z$ axis [Fig.~\ref{fig:trion}(c)]. As a result, even
after the trions have vanished (${\rm e}^{ - t / \tau^T } \to 0$),
the electron spin polarization is nonzero and it oscillates with
frequency $\Omega$. In Eq.~(\ref{sta}), the $\eta$-independent
term describes the spin precession of resident electrons left
after the pump pulse ends while the terms proportional to $\eta$
take into account the electrons left after the trions have
recombined and decayed exponentially. At the moment of trion
recombination, these electrons become polarized antiparallel to
$z$, their spin is added to the total electron spin and rotates as
well with frequency $\Omega$ in the $(y,z)$ plane.

\begin{figure}[htb]
\centering
\includegraphics[width=1\linewidth]{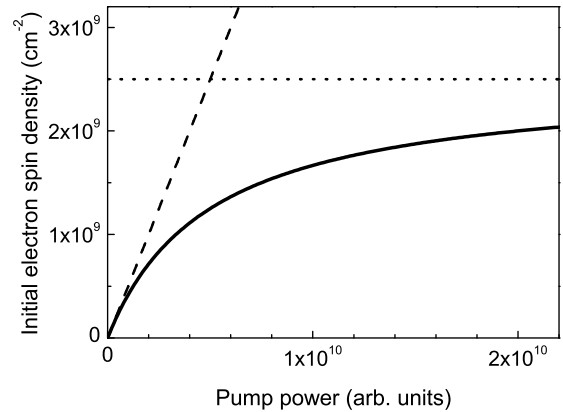}
\caption{Initial spin of the 2DEG as function of the pumping power
(solid). The asymptotic lines correspond to linear growth (dashed)
and saturation (dotted). Here it has assumed that the laser is
circular polarized and is resonant with the trion energy. The pump
power is given in units of photocreated trions in the linear
regime.} \label{fig:satur:theor}
\end{figure}

Time-resolved KR can originate both from the electron and the
trion spin polarizations. According to Eq.~(\ref{sza}) the
electron contribution to the Kerr signal contains both exponential
monotonous and oscillatory components, whereas the trion
contribution shows only a monotonous behavior, proportional to
${\rm e}^{ - t/\tau^T}$. It should be stressed that the initial
number of photogenerated trions under resonant excitation,
$n_0^T$, cannot exceed $n_e/2$, where $n_e$ is the density of the
2DEG. Thus, the factor $n_0^T$ in Eq.~(\ref{sza}) increases
linearly with the pump intensity for small excitation density and
then saturates with density increase at the value $n_e/2$. In the
simplest model
\begin{equation}\label{trion_res_sat}
n_0^T= \frac{n_e}{2}
G\tau_0^T/(1+G\tau_0^T),
\end{equation}
where $G$ is the generation rate being proportional to the pump
power. Thus the initial spin polarization of the 2DEG shows a
saturation behavior, see Fig.~\ref{fig:satur:theor} which can be
related to saturation of trion absorbtion.

\subsection{Resonant excitation of excitons in a diluted
2DEG}\label{exciton}

If the pump photon energy is tuned to the exciton transition then,
at low temperates where $k_B T < E^T_B$ (with the Boltzmann
constant $k_B$, and the trion binding energy $E^T_B$), the
photogenerated excitons tend to bind into trions as long as they
find resident electrons with proper spin orientation. The trion
thermal dissociation, on the other hand, can be neglected.
Experimentally, the exciton contribution to the Kerr oscillating
signal is determined by the Larmor spin precession of the
electrons in the excitons. In the pump-probe experiment the
correlation between the electron and hole spins via the
electron-hole exchange interaction is suppressed, see
Refs.~[\onlinecite{toulouse,toulouse1}]. Therefore, the spin ${\bm
s}^X$ of an electron in an exciton precesses about an in-plane
magnetic field with the same frequency as that of a resident
electron. Note that if the heavy holes forming excitons are
unpolarized the excitons can be labelled by the electron spin
${\bm s}^X$. Moreover, the trions formed from these excitons are
unpolarized as well.

At zero magnetic field the rates of an electron and an exciton
binding into a trion are given by
\begin{equation} \label{second}
\left( \frac{dn_{\pm}}{dt} \right)_{\rm T} = \left(
\frac{dn^X_{\mp}}{dt} \right)_{\rm T}= - \gamma n_{\pm} n^X_{\mp}
\:,
\end{equation}
where $n_{\pm}, n^X_{\pm}$ are the densities of electrons and
excitons with electron spin $\pm 1/2$, and $\gamma$ is a constant.
In addition to this constant the system is characterized by four
times, namely, the exciton and trion radiative lifetimes,
$\tau^X_0$ and $\tau^T_0$, respectively, and the spin relaxation
times of resident electrons ($\tau_s$) and excitons ($\tau_s^X$).
In the presence of an in-plane magnetic field ${\bm B} \perp z$,
the spins of the resident electrons and the electrons in excitons
precess with the same frequency $\Omega$. In the coordinate frame
which rotates around ${\bm B}$ at a rate $\Omega$, one can apply
the simple equation (\ref{second}) to describe the spin-dependent
decay of resident electrons and photoexcited excitons.

At low excitation intensities satisfying the condition $n^X_0 \ll
n_e/2$ (here $n^X_0$ is the number of photogenerated excitons) the
total spin of the electron gas after the decay of all excitons can
be estimated as
\begin{equation}\label{se_low_taus}
|{\bm S}_e| = \frac{\tilde \tau}{2\tau_b}n^X_0,
\end{equation}
where $\tilde \tau$ is the total lifetime of the exciton spin,
including the radiative decay time, the time of exciton binding
into a trion, $\tau_b \sim (\gamma n_e)^{-1}$, and the spin
relaxation time $\tilde{\tau}^{-1} = (\tau_0^X)^{-1} + \tau_b^{-1}
+ (\tau_s^X)^{-1}$. In order to simplify the analysis we assume
the time of exciton binding into a trion, $\tau_b \sim (\gamma
n_e)^{-1}$, to be shorter than the exciton radiative lifetime
$\tau^X_0$ and the spin relaxation time of the electron in an
exciton $\tau^X_s$. In this case, shortly after the pulsed optical
excitation all excitons are bound to trions and the QW structure
contains $n^X_0$ trions and $n_e - n^X_0$ resident electrons with
a total precessing spin
\begin{equation} \label{se_low}
|{\bm S}_e| = n^X_0/2,    \quad n_0^X\leq n_e/2.
\end{equation}
As a result, $n^X_0$ spins of resident electrons contribute to the
Kerr rotation oscillations.

At higher excitation intensity, $n^X_0 \geq n_e/2$, and the
spin-polarized excitons extract almost at once $n_e/2$ electrons
to form trions. Therefore, in absence of electron-in-exciton spin
relaxation processes the trion density cannot exceed $n_e/2$, thus
the total spin density of the electron gas is limited by $n_e/4$.
The electron-in-exciton spin relaxation allows to convert the
remaining $n^X_0 - (n_e/2)$ excitons in trions. Obviously, the
maximum number of formed trions can not exceed the concentration
of background electrons, $n_e$.  The total spin of resident
electrons after the excitons and trions have recombined can be
estimated as (provided that the holes are unpolarized)
\begin{equation}\label{se_exc}
|{\bm S}_e|  \approx \frac{1}{4} \left\{
\begin{array}{cc}
n_e - \frac{2 n^X_0 - n_e}{1 + (2 \tau^X_s/\tau_0^X)} , & n_e
> \frac{2n^X_0 - n_e}{1 + (2 \tau^X_s/\tau_0^X)} \\
0, & \mbox{otherwise}.
\end{array}
\right.
\end{equation}
This equation is valid both for $B=0$ and $B \neq 0$ when $n^X_0
\geq n_e/2$, otherwise Eq. \eqref{se_low} holds. At $n^X_0 =
n_e/2$, the values of $|{\bm S}_e|$ given by Eq.~(\ref{se_low})
and Eq.~(\ref{se_exc}) coincide and are equal to $n_e/4$.

\begin{figure}[htb]
\centering
\includegraphics[width=1\linewidth]{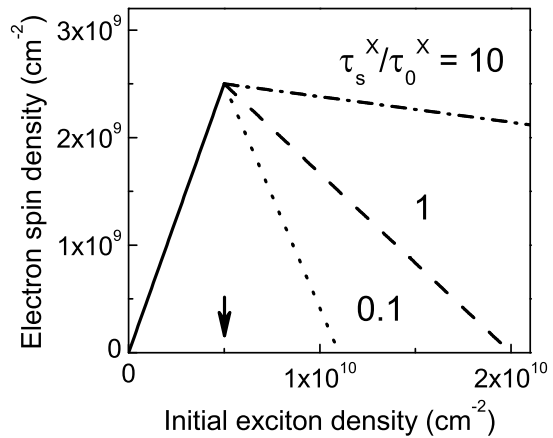}
\caption{Schematic plot of the spin of the 2DEG after exciton and
trion recombination as function of the initial exciton density
(i.e. pumping power). Solid curve is plotted according to Eq.
\eqref{se_low} and dotted, dashed and dash-dotted curves are
plotted according to Eq. \eqref{se_exc}. These curves correspond
to different values of $\tau_s^X/\tau_0^X=0.1$, $1$ and $10$,
respectively. $n_e=10^{10}$~cm$^{-2}$. The cross-over density
$n_0^X=n_e/2$ is shown by the arrow.} \label{fig:excexc:theor}
\end{figure}

When deriving the above equation we have neglected the spin
relaxation of the resident electrons assuming $\tau^X_s \ll
\tau_s$. In experiment the exciton radiative lifetime, $\tau^X_0$,
may be comparable with the trion formation time, $\tau_b$. In this
case the above estimation should be taken as a qualitative result
predicting a non-monotonous dependence of the Kerr signal
amplitude on pump intensity. This non-monotonous behavior is
illustrated in Fig.~\ref{fig:excexc:theor}. An initial linear
growth of $|\bm S_e|$ followed by a linear decrease is seen. The
decrease of initial electron spin as function of pump intensity is
steeper for smaller values of $\tau_s^X/\tau_0^X$, i.e. for
shorter hole spin relaxation times. It is worth to stress that in
this regime the electron spin polarization vanishes at very high
pumping whereas, under resonant trion excitation, $|{\bm S}_e|$
monotonously increases with increasing pump power and saturates at
$n_e/4$.

\subsubsection*{Detection aspects}

Turning to the detection aspects, we find that selective
addressing of the exciton resonance also results in temporal
oscillations of the probe-pulse Kerr rotation. The modulation
comes from the photoinduced difference in the resonance
frequencies $\omega_{0,\pm}$ and/or the non-radiative damping
rates $\Gamma_{\pm}$. Both $\omega_{0,+} - \omega_{0, -}$ and
$\Gamma_{+} - \Gamma_{-}$ become nonzero taking into account the
exchange interaction between an electron in an exciton and the
resident electrons of the 2DEG: the first difference is related to
the Hartree-Fock renormalization of the electron energy in the
spin-polarized electron gas, and the second one is related to the
spin dependence of the electron-exciton scattering~\cite{Ast00}.
As a result, rotation of the total spin of the electron ensemble
leads to a modulation of the exciton resonance frequency and
non-radiative broadening and, thus, to oscillations of the KR
angle. We note that an in-plane magnetic field results also in
spin precession of the electron in an exciton and the total Kerr
signal will be a superposition of 2DEG and exciton signals.

The situation for the probe tuned to the trion resonance is
qualitatively the same. The KR amplitude will contain components
arising due to spin precession of the 2DEG and to the
electron-in-exciton spin precession. However, it is expected that
detection at trion resonance will be less sensitive to the exciton
spin dynamics as compared to detection at the exciton resonance,
see Sec.~III.B. We note here that the amplitude of the KR signal
induced by the same number of coherent electron spins will be
different for detection at the trion or the exciton energies. The
difference comes from the different oscillator strengths of these
resonances \cite{Ast00} and from the different efficiencies of
signal modulation.

\subsection{Resonant excitation of excitons in a dense 2DEG}

An increase of the 2DEG density and its Fermi level leads to
dissociation of trions due to state-filling and screening effects.
We consider an intermediate situation where the trions are
suppressed, $n_ea_T^2>1$, but the excitons are not, $n_e a_B^2 \ll
1$, where $a_B$ is the exciton Bohr radius and $a_T$ is the
characteristic trion radius. Therefore, the scenario of spin
coherence generation for electrons involves two subsystems, the
2DEG and the spin-polarized excitons resonantly excited by the
circular polarized optical pulse. The electrons being initially
unpolarized can gain spin polarization due to the
electron-electron exchange interaction in electron-exciton
flip-flop scattering processes. At zero magnetic field, the
flip-flop scattering rates for the resident electrons and those
bound in excitons are given by
\begin{equation} \label{nn'}
\left( \frac{dn_{\pm}}{dt} \right)_{\rm exch} = \left(
\frac{dn^X_{\mp}}{dt} \right)_{\rm exch}= - \gamma' (n_{\pm}
n^X_{\mp} - n_{\mp} n^X_{\pm})\:,
\end{equation}
where $\gamma'$ is a constant different from $\gamma$ in
Eq.~(\ref{second}). The total number of electrons, $n_e = n_+ +
n_-$, is fixed while the exciton density $n^X = n^X_+ + n^X_-$
decays to zero.

Similarly to the scenario in Sec.~III.C, we assume that the holes
in the excitons are unpolarized and the spins of both the resident
and bound electrons precess with the same frequency. The Kerr
rotation signal is a result of the difference in shift between the
Fermi levels for spin-up and spin-down electrons and the
concomitant renormalization of the resonance frequencies
$\omega_{0,\pm}$ due to the exchange interaction between the
spin-polarized carriers.

At weak pump intensities so that
\[
\gamma n^X_0 \ll \frac{1}{\tilde{\tau}} = \frac{1}{\tau^X_0} +
\frac{1}{\tau^X_s} + \gamma n_e\:,
\]
and for negligible resident-electron spin relaxation,
$\tilde{\tau} \ll \tau_s$, one can use, in agreement with Eq.
\eqref{se_low_taus}, the following estimate for the amount of
resident-electron spins oriented by excitons: $ |{\bm S}_e| =
\frac12 \gamma \tilde{\tau} n_e n^X_0$. With increasing pump
intensity the electron spin saturates at $|{\bm S}_e|_{\rm max}
\sim \frac12 n_e [1 - (\tilde{\tau}/\tau^X_s)]$ when all 2DEG
electrons have become spin-polarized.

It should be noted that a similar description can be applied for
resonant photoexcitation of excitons in the high-temperature
regime where $k_B T \gg E_B^T$, but $k_B T < E_B^X$, so that the
trion states are thermally unstable but the exciton bound states
are still stable.

\subsection{Nonresonant excitation of carriers}

In the case of non-resonant pumping, the photogeneration of free
electrons and holes is followed by their separate, uncorrelated
energy relaxation toward the bottoms of the corresponding bands.
The energy relaxation is accompanied by carrier spin relaxation.
The holes have lost their spin after few scattering events. The
total spin of the electron ensemble after energy relaxation can be
estimated as
\[
S_z \approx \frac{\tau_s^*}{\tau_s^* + \tau_\epsilon} n^e_0,
\]
where $n^e_0$ is the number of photogenerated electrons,
$\tau_\epsilon$ is the energy relaxation time and $\tau_s^*$ is
the spin relaxation time of hot electrons. We note that $\tau_s^*$
can be much shorter than the spin coherence time $\tau_s$ of the
electron gas in quasi-equilibrium, entering Eq. \eqref{1}.

After the particles have reached the band bottoms they can bind to
form excitons and trions. In the diluted electron gas and for
moderate pumping densities, the trion formation is more preferable
and the subsequent spin dynamics can be described by the model in
Sec.~II.B. At very strong pumping, when the number of
photogenerated electrons exceeds that of the resident electrons,
or for a dense 2DEG, for which the trions are suppressed,
formation of excitons takes place. The spin dynamics in this case
can be described by the scenario in Sec.~II.D.

\subsection{Summary}

We have developed a comprehensive model to describe the Kerr
rotation signal in QWs with varying 2DEG densities and for
different excitation regimes. The signal dynamics reveals spin
oscillations of the resident electrons and/or of the electrons
forming excitons. The pump power dependence of the Kerr rotation
amplitude is as follows: at low pumping the amplitude grows
linearly with the pump power, whereas in the limit of high pumping
the amplitude behavior strongly depends on the excitation energy.
For trion resonant excitation it saturates (as the number of
generated trions is limited by half of the resident electron
concentration), while for resonant excitation of excitons (at $k_B
T < E_B^T$ and in the diluted 2DEG) the amplitude exhibits a
non-monotonic behavior and eventually vanishes.

A theoretical model of excitation of electron spin polarization by
short laser pulses tuned to the trion resonance has been also
proposed in Refs.~[\onlinecite{Sha03,Ken06,Gre06}]. Their approach
is based on analysis of the temporal dynamics of the coherent
quantum beats between an electron and a trion localized by a
quantum dot or by well width fluctuations in a QW plane.

Our approach and the approach in
Refs.~[\onlinecite{Sha03,Ken06,Gre06}] are essentially equivalent.
Indeed, let us consider a localized unpolarized electron. Its spin
density matrix is diagonal with $\rho_{1/2,1/2} = \rho_{-1/2,-1/2}
= 1/2$. After optical excitation with a $\sigma_+$
circular-polarized pulse a superposition state of the localized
electron and a trion is formed. After the trion radiative decay
the electron remains, but its density matrix is different from the
initial one. For example, in the limiting case where the hole spin
relaxation time is much shorter than the trion radiative lifetime
which, in turn, is shorter than the electron spin relaxation time,
the electron spin density matrix after trion recombination remains
diagonal with
\begin{equation}
\label{A6}  \rho_{1/2,1/2} = \frac12 \left( 1 - \frac{|D|^2}{2}
\right),\: \rho_{-1/2,-1/2} = \frac12 \left( 1 + \frac{|D|^2}{2}
\right)\:.
\end{equation}
Here $D$ is a complex coefficient which depends on the optical
pulse parameters. The resident electron acquires spin-down
polarization under pulsed $\sigma^+$ photoexcitation and a train
of such pulses leads to complete spin polarization. The same
single electron spin density matrix, Eq. \eqref{A6}, describes an
ensemble where the number of spin-up electrons is smaller than
that of spin-down electrons. The constant $D$ in this case has a
transparent physical sense: it characterizes the efficiency of
singlet trion formation under polarized excitation of the
unpolarized ensemble. One can readily check that this result is
equivalent to Eq. \eqref{st0} taken at $t=0$.

\section{Experimental results}

\subsection{Experimentals}

The studied CdTe/Cd$_{0.78}$Mg$_{0.22}$Te QW heterostructure was
grown by molecular-bean epitaxy on an $(100)$-oriented GaAs
substrate followed by a 2 $\mu$m CdTe buffer layer. It has 5
periods, each of them consisting of a 110 nm thick
Cd$_{0.78}$Mg$_{0.22}$Te barrier and a 20 nm thick CdTe QW. An
additional 110~nm~thick barrier was grown on top of this layer
sequence to reduce the contribution of surface charges. The
barriers contain 15~nm~layers doped by In donors. Undoped
15~nm~spacers separate the modulation-doped layers from the QWs.
Electrons from the barrier donors, being collected into QWs,
provide there a 2DEG with a low density of about $n_e= 1.1\times
10^{10}$~cm$^{-2}$. We have observed very similar experimental
results for structures with single CdTe/Cd$_{0.7}$Mg$_{0.3}$Te QWs
of 12~nm and 8~nm widths, respectively. This confirms the general
character of the reported data.

A time-resolved pump-probe Kerr rotation technique was used to
study the coherent spin dynamics of the electrons \cite{Aws02}. We
used a Ti:Sapphire laser generating 1.5 ps pulses at a repetition
frequency of 75.6 MHz. The laser beam was split in pump and probe
beams and the time delay between the pump and probe pulses was
varied by a mechanical delay line. The pump beam was circular
polarized by means of an elasto-optical modulator operated at 50
kHz. The probe beam was linearly polarized, and rotation of the
polarization plane was measured by a balanced photodetector. The
time-resolved Kerr rotation signal allows to follow the evolution
of the spin coherence of carriers and their complexes generated by
the pump pulses. From an analysis of the decay of the Kerr
rotation amplitude the spin dephasing time of the electron
ensemble $T_2^*$ can be extracted. The details of this analysis
through which the mechanisms of spin dephasing of the 2DEG can be
understood have been reported elsewhere \cite{Zhu06a, Yak07}.

The Kerr rotation technique has been used in two regimes. For
degenerate Kerr rotation the pump and probe beams have the same
photon energy, as they originate from the same laser. This regime
will be denoted as one-color experiment here. We performed,
however, also a two-color experiment where the pump and probe
energy can be tuned independently. For that purpose two
synchronized Ti:Sapphire lasers were used. Experiments were
performed in magnetic fields (0 - 7 T) applied in the plane of the
structure, i.e. in the Voigt geometry. The sample temperature was
tuned in a range 1.9 - 100 K.

Further, time-resolved photoluminescence was used to study
recombination dynamics of excitons and trions. The same laser as
described above was used for excitation, and time resolved
emission spectra have been recorded by a synchro-scan streak
camera connected to a 0.5 meter spectrometer. The time resolution
in this experiment was about 5 ps.

\begin{figure}[htb]
\centering
\includegraphics[width=1\linewidth]{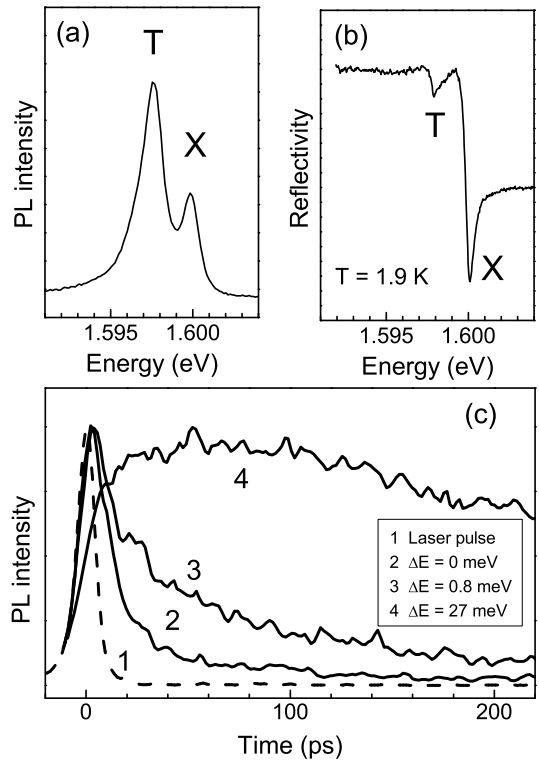}
\caption{(a) Photoluminescence spectrum of a 200~\AA~
 CdTe/Cd$_{0.78}$Mg$_{0.22}$Te QW measured under nonresonant cw
excitation with a photon energy of 2.33 eV. The exciton (X) and
trion (T) resonances are separated by 2~meV, which is the trion
binding energy. (b) Reflectivity spectrum of the same structure.
The oscillator strength of the exciton resonance is ten times
larger than the trion one. (c) Kinetics of the photoluminescence
measured by a streak camera under 1.5~ps pulsed excitation
resonant to the trion energy (curve 2) and detuned from it by
0.8~meV and 27 meV to higher energies. The dashed line shows the
laser pulse. }\label{fig:A2}
\end{figure}

A typical photoluminescence (PL) spectrum of the QW measured at a
temperature $T = 1.9$~K is shown in Fig.~\ref{fig:A2}~(a). It
shows the exciton and trion recombination lines separated by
2~meV, corresponding to the trion binding energy. The full width
at half maximum of the exciton line is about 0.5~meV and is mainly
due to exciton localization on the QW width fluctuations.

The reflectivity spectrum of the same QW in the energy range of
the exciton and trion resonances is given in
Fig.~\ref{fig:A2}~(b). Following the procedure described in
Ref.~[\onlinecite{Ast00}] we have evaluated the oscillator
strengths of the resonances and found that the exciton oscillator
strength is ten times larger than that of the trion resonance.
This fact should be taken into account when the intensity of the
Kerr rotation signals measured at the exciton and trion energies
are compared: Firstly, the probe response is proportional to the
oscillator strength and, secondly, the number of photogenerated
carriers is proportional to it.

For interpretation of the results of the spin dynamics
experiments, information on the recombination dynamics of excitons
and trions under resonant excitation is necessary. We performed
corresponding measurements under linearly polarized excitation
using the streak camera for detection. The results for pumping
into the exciton and trion resonances are very similar to each
other. The typical recombination kinetics for the trion is given
in Fig.~\ref{fig:A2}~(c). For resonant excitation at the trion
energy (curve 2, detuning $\Delta E=0$) about 80$\%$ of the PL
intensity decays with a time of 30 ps and the rest decays with a
time of 100~ps. When the excitation energy is detuned by 0.8~meV
above the trion resonance a redistribution of the two exponential
decays with 30 and 100~ps time occurs in favor of the longer decay
component. Such a behavior is typical for QW emission \cite{Ciu00,
Yak07}. The shorter decay time, 30~ps, can be attributed to the
radiative recombination time of trions and excitons generated in
the radiative cone where their wave vectors is transferred to the
photons. For excitons scattered out of the radiative cone, a
longer time of about 100 ps is required to be returned to the cone
via emission of acoustical phonons. The exciton luminescence
lifetime is slowed down to 100~ps in this case. The recombination
of trions is not restricted to the radiative cone conditions as
the electron left in the system can take the finite momentum to
satisfy the wave vector conservation. Therefore we may expect the
fast decay of trion PL of about 30 ps even for nonresonant
photoexcitation. In most cases, however, for these experimental
conditions the trions are formed from photogenerated excitons,
which dominate the absorption due to their larger oscillator
strength. As a result the decay of the trion PL in this case is
determined not by trion recombination but by trion formation, and
will monitor the exciton lifetime. For small detuning exemplified
by curve 3, the fast 30 ps process coexists with the longer (100
ps) one. When the excitation energy is tuned to the band-to-band
absorption (curve 4, $\Delta E=27$~meV), the PL decay is extended
to 250~ps as additional time is required for the free carriers to
be bound to excitons. Note that the exciton binding energy in the
studied QW is $E_B^X=12$~meV.

Below we present the experimental results for the two-color and
the degenerate pump-probe experiment. Further, we show the
experimental data for the pump power, temperature and magnetic
field dependencies of the Kerr rotation signals.

\subsection{Two-color pump-probe}

The two-color Kerr rotation technique enables independent tuning
of the energies of the pump and probe beams. This allows to
perform experiments with either constant excitation or detection
conditions, which simplifies identification of the studied
relaxation processes.

\begin{figure}[htb]
\centering
\includegraphics[width=1.1\linewidth]{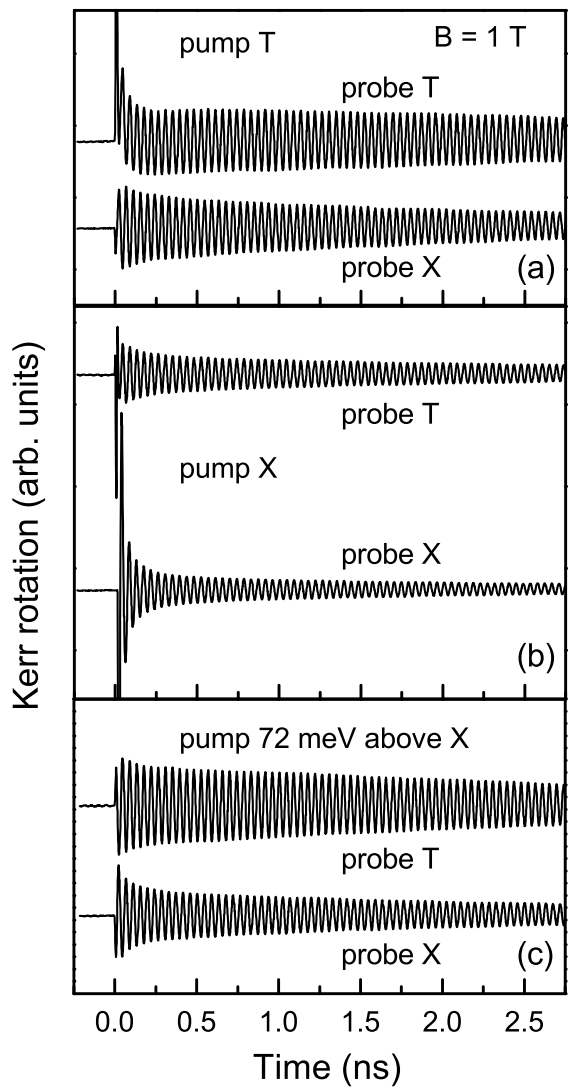}
\caption{KR signals measured by a two-color technique at
$T=1.9$~K, $B=1$~T and various pump excitation energies: (a)
resonant with trion at 1.5982 eV, (b) resonant with exciton at
1.6005 eV, and (c) nonresonant at 1.6718 eV, which is 72 meV above
the exciton resonance. Pump density 56~W$/$cm$^2$ and probe
density 8~W$/$cm$^2$.}\label{fig:A3}
\end{figure}

Figure~\ref{fig:A3} shows Kerr rotation signals detected at the
trion and exciton resonances for three pump energies: (a) resonant
with the trion, (b) resonant with the exciton, and (c)
non-resonantly excited 72 meV above the exciton energy. The sample
is subject to an in-plane magnetic field $B=1$~T. All signals show
damped oscillations with a frequency of 23 GHz, which is the
Larmor precession frequency of the electron spins. This frequency
corresponds to the Zeeman splitting of the conduction band
electrons with a $g$ factor value of 1.64 which is in good agreement
with literature data for the electron $g$ factor in CdTe/(Cd,Mg)Te
QWs \cite{Sir97}. Another common feature of all signals shown in
Fig.~\ref{fig:A3} is the appearance of long-living spin beats
which are observed beyond delays of 2.7 ns. The typical
recombination times of excitons and trions do not exceed 30-100~ps
for resonant and quasi-resonant excitation and 250~ps for
nonresonant excitation [see Fig.~\ref{fig:A2}~(c) and Sec. III.
A.], therefore we identify the long-living signal with the
coherent spin precession of the resident electrons in the 2DEG.
One can see that this coherence is excited efficiently for all
pump energies and can be detected by probing both the trion and
exciton resonances.

Some of the signals shown in Fig.~\ref{fig:A3} contain a
short-living part right after the pump pulse with a typical decay
time of 50-70~ps. This part is especially pronounced for the
``pump X/probe X'' condition (i.e. pump and probe are degenerate
with the exciton resonance), see panel (b). This fast component is
related to the exciton contribution to the Kerr rotation signal,
see Sec. II. C. and discussion below. To extract the times and
relative amplitudes of the short and long-living components in the
spin beat signals each trace has been fitted with a bi-exponential
decay function:
\begin{equation} \label{st100}
y(t) = (A{\rm e}^{ - t/\tau_1} +  B{\rm e}^{ - t/\tau_2})
\sin(\omega t + \varphi) ,
\end{equation}
where $A$ and $B$ are constants describing the amplitudes of the
fast ($\tau_1$) and slow ($\tau_2$) components, respectively,
$\omega$ is the Larmor frequency which is taken to be the same for
both components, and $\varphi$ is the initial phase.

The parameters extracted from these fits are collected in Table 1.
All signals, except the one for ``pump T/probe T'', are symmetric
with respect to the abscissa. The ``pump T/probe T'' signal shows
an initial relaxation of the centre-of-gravity of the electron
beats with a time constant of about 75~ps, which can be attributed
to the spin relaxation time of the hole in the trions, see the
second term in Eq. \eqref{sza} and Ref. \cite{Zhu06a}. A detailed
analysis of hole spin coherence in CdTe/(Cd,Mg)Te QWs will be
reported elsewhere.

\begin{table}[htbp]
\caption{Decay times $\tau_1, \tau_2$  and amplitude ratios $A /
 B$ extracted from bi-exponential fits to the experimental data using Eq. \eqref{st100} in Fig.~\ref{fig:A3}. $B=1$~T and
 $T=1.9$~K.}
 \begin{tabular}{c|c|c|c}
         & pump T       & pump X        & nonresonant \\
\hline
probe T  & -/5.7~ns     & 40~ps/3.5~ns  & 56~ps/3.6~ns \\
         & 0/1          & 0.5/0.5       & 0.2/0.8 \\
\hline
probe X  & -/2.6~ns     & 50~ps/2.0~ns  & 70~ps/2.8~ns \\
         & 0/1          & 0.9/0.1       & 0.5/0.5
 \end{tabular}
\end{table}

The decay times and relative amplitudes in Table 1 are given for
$B=1$~T and $T=1.9$~K. It is important to note that these
parameters depend strongly on pump intensity, magnetic field
strength and lattice temperature. These dependencies and the
underlying physical mechanisms will be discussed in detail below.
Here we focus our attention on the pump energy dependence of these
parameters. We first concentrate on the relative amplitudes as
they are a key to understanding the generation of 2DEG spin
coherence and the role of the trions in this process.

For the ``pump T/probe T'' regime only the long-living electron
signal of a 2DEG is observed [see panel (a) of Fig.~\ref{fig:A3}].
This is in line with the model expectations as in this case only
trions are photogenerated. As these trions are in the singlet
ground state with antiparallel electron spins, they do not
contribute to the KR signal. Moving the pump energy in resonance
with the exciton and further to a band-band transition leads to
appearance of the fast decaying component for the signal probed at
the trion energy. This can be attributed to the spin dynamics of
the exciton which is excited either resonantly or non-resonantly,
see Sec. II C, E. The probe has a finite spectral width (about 1
meV) and tuned to the trion resonance slightly overlaps with the
exciton resonance. The shortening of the electron spin dephasing
time from 5.7 down to 3.5-3.6 ns is attributed to heating of the
2DEG by photocarriers with excess kinetic energy (for a detailed
discussion see sections III.D and III.E).

The Kerr rotation signal probed at the exciton energy has two
contributions: (i) given by the coherent precession of the
electrons in excitons and (ii) given by the spin precession of the
2DEG. The former decays with the exciton recombination time. This
fast exciton component is clearly seen in panels (b, c) and is
absent for pumping at the trion resonance. For nonresonant
excitation its relative amplitude does not exceed 50\%.  In this
case electrons and holes are photogenerated 72 meV above the
exciton resonance, therefore they have a high probability to
scatter and relax separately to the bottoms of their bands, where
they are bound into trions and excitons. The relative amplitudes
of the fast and long-lived signals reflect the probability of
trion and exciton formation. It may be expected that trion
formation is preferable because the 2DEG density exceeds by at
least an order of magnitude the concentration of photocarriers
which is in line with the experimental findings.

A very different ratio of the relative amplitudes, 90\% for the
fast decay and 10\% for the long-living dynamics, is seen for the
signal when resonantly pumped at the exciton energy and detected
at the exciton [see panel (b) in Fig.~\ref{fig:A3}]. There are at
least two factors which favor an exciton population in comparison
with a trion one under resonant pumping of the excitons. First,
the photogeneration leads to formation of excitons with very low
kinetic energy, therefore they remain within the radiative cone
and quickly recombine (during 30-50 ps) prior to becoming bound to
trions \cite{Ciu00}. Second, a part of excitons is localized, so
that they are not mobile and can not reach the sites in the QW
where the background electrons are localized. Consequently, the
formation of trions out of this exciton reservoir is suppressed.
Moreover, the ratio between the contributions of the excitons and
the 2DEG to the KR signal is spectrally dependent: detection at
the exciton resonance is more sensitive to the spin precession of
the electron in the exciton, see Sec. II. C.

The results of the two-color experiments presented in
Fig.~\ref{fig:A3} are in good agreement with the model
expectations, namely: (i) the signal oscillating with the electron
Larmor frequency is contributed by resident electrons of the 2DEG
and by electrons precessing in the excitons, (ii) trion formation
resulting either from resonant photoexcitation of the trions or
from capture of excitons or free carriers is a very efficient
mechanism for spin coherence generation in a diluted 2DEG.

\subsection{Degenerate pump-probe Kerr rotation}

The degenerate pump-probe technique is simpler in technical
realization and therefore is a more common method to address spin
coherence. In this case the pump and probe pulses are generated by
the same laser beam without additional spectral selection.
Examples of degenerate Kerr signals can be found already in
Fig.~\ref{fig:A3}, for example the ``pump T/probe T'' and ``pump
X/probe X'' traces measured for coinciding energies of the two
lasers.

\begin{figure}[htb]
\centering
\includegraphics[width=1\linewidth]{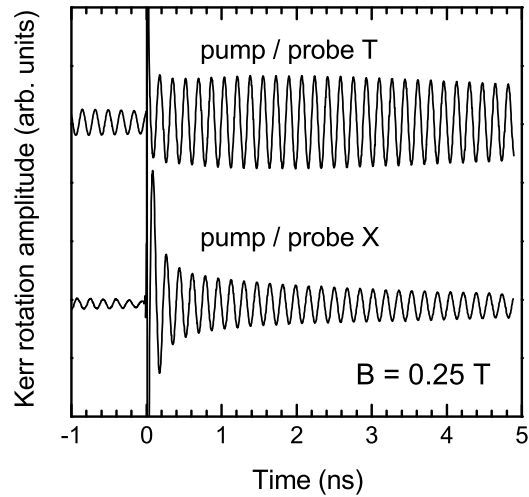}
\caption{Kerr rotation measured by degenerate pump-probe resonant
either with the trion or the exciton energies. $T=1.9$~K, pump
density is $1.5$~W$/$cm$^2$ and probe density is
$0.3$~W$/$cm$^2$.}\label{fig:exp2}
\end{figure}

The degenerate pump-probe signal presented in Fig.~\ref{fig:exp2}
was measured with the use of one laser, as will be the case for
the most experiments presented in the rest of this paper. It is
consistent with spin dynamics studied by two-color technique.
Namely, only long-lived spin beats of resident electrons are
observed when the laser is resonant with the trion and two
component decay is seen for the laser hitting exciton. The
experimental conditions were modified as compared with
Fig.~\ref{fig:A3} in order to achieve longer electron spin
coherence times. Namely, the pump density was reduced to
$1.5$~W$/$cm$^2$ and a weaker magnetic field $B= 0.25$~T was
applied~\cite{Zhu06a}. The delay time range between pump and probe
in Fig.~\ref{fig:exp2} covers 6 ns. Under these conditions the
spin dephasing time of the resident electrons reaches 13.7 ns for
pumping in the trion resonance and 4.2 ns for pumping in the
exciton resonance. Moreover the spin coherence does not fully
decay during the time interval of 13.2 ns between the pump pulses
as is clearly seen by the beats at negative delays in
Fig.~\ref{fig:exp2}.

\subsection{Dependence on pump density}

In this part we investigate the modifications of Kerr rotation
signal occurring when the pump density is increased in a wide
range of pump powers $P$ from 1 to 320~W$/$cm$^2$. Degenerate
pump-probe resonant either with the exciton or the trion
transition energies is used.

\begin{figure}[htb]
\centering
\includegraphics[width=1.1\linewidth]{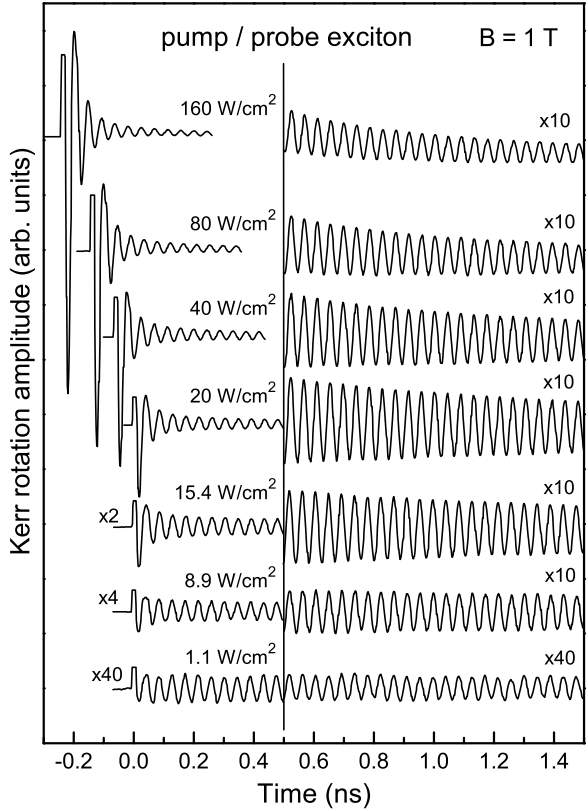}
\caption{Kerr rotation signals for degenerate pump-probe resonant
with the exciton energy measured at different pump densities.
$T=1.9$~K. For clarity of presentation up-scaling factors have
been used for delays exceeding 0.5 ns. Also the initial parts of
the signals for pump densities exceeding 20~W$/$cm$^2$ have been
shifted in time to negative delays. For all signals a coherent
signal at zero delay formed due to the temporary overlap of the
probe with the pump has been removed.}\label{fig:exp3}
\end{figure}

Results for excitation at the exciton transition energy are
collected in Fig.~\ref{fig:exp3}. With increase of the pump
density a very different behavior is observed for the
fast-decaying exciton component and the long-living 2DEG
component. One can clearly see that the exciton part is rather
weak at low powers, evidencing the high probability for an exciton
to become bound to a resident electron and form a trion. At higher
excitation power the fast-decaying component becomes more
pronounced and even dominates in the power range $P=20\ldots 180$
W$/$cm$^2$ where the concentration of photogenerated excitons
exceeds the 2DEG density. Moreover the absolute amplitude of the
2DEG signal at longer delays is decreased for pump powers
exceeding 20~W$/$cm$^2$. This result will be discussed below
together with the data plotted in Fig.~\ref{fig:exp3}.

\begin{figure}[htb]
\centering
\includegraphics[width=1\linewidth]{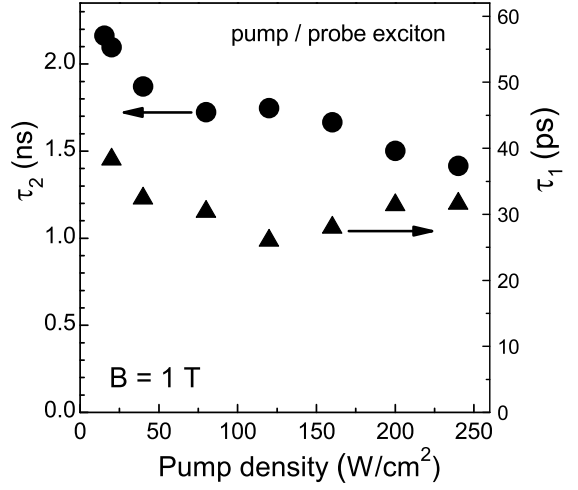}
\caption{Pump density dependence of the 2DEG spin dephasing time
$T_2^* = \tau_2$ (circles) and the time of the fast decay $\tau_1$
(triangles) related to the exciton recombination dynamics.
$T=1.9$~K. }\label{fig:exp3t}
\end{figure}

Relaxation times obtained from the fit with Eq.(\ref{st100}) are
given in Fig.~\ref{fig:exp3t}. The fast decaying component falls
in the range of 25-40~ps and is independent of the pump density.
It is clearly determined by the radiative recombination of
excitons resonantly excited in the light cone. The decay time of
the long-living component related to the spin dephasing of the
2DEG decreases from 2.1~ns at 15~W$/$cm$^2$ down to 1.4~ns at
240~W$/$cm$^2$.  A possible reason for this behavior can be the
heating of the resident electrons by the photoexcitation, leading
to a reduction of $T_2^*$.

\begin{figure}[htb]
\centering
\includegraphics[width=1.1\linewidth]{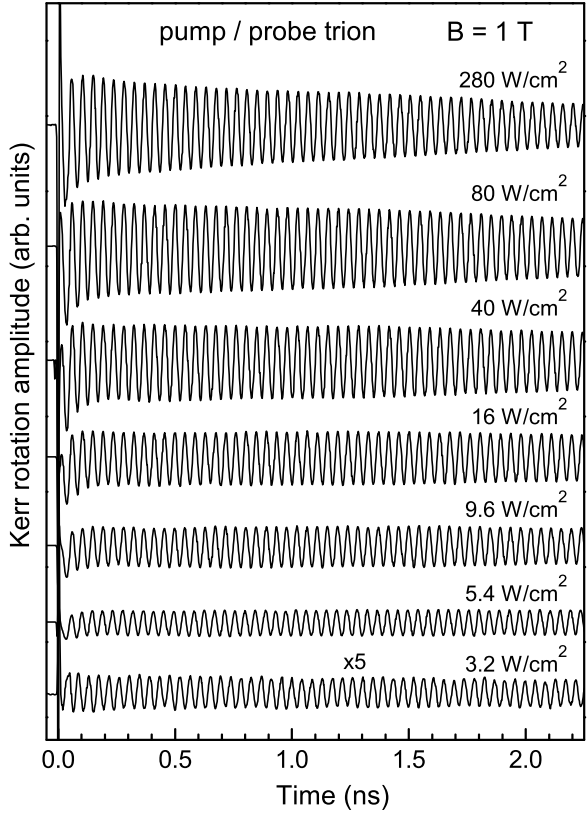}
\caption{Kerr rotation signals for degenerate pump-probe resonant
with the trion energy measured at different pump densities.
$T=1.9$~K.}\label{fig:exp4}
\end{figure}

\begin{figure}[htb]
\centering
\includegraphics[width=1.1\linewidth]{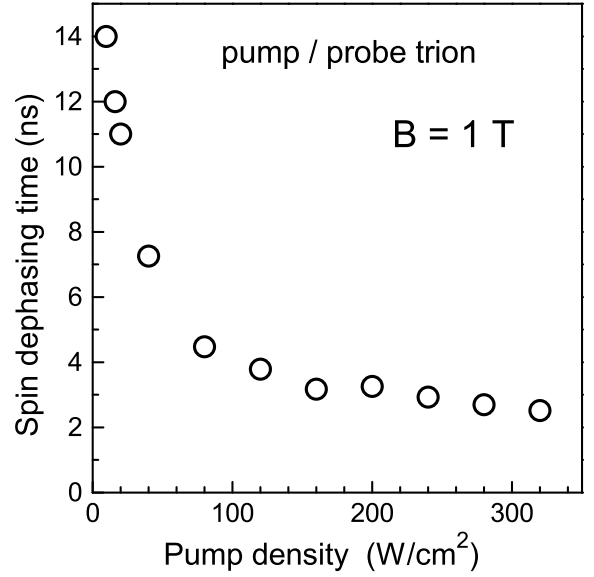}
\caption{Pump density dependence of the 2DEG spin dephasing time
$T_2^*$. $T=1.9$~K. }\label{fig:exp4t}
\end{figure}

Kerr rotation signals for resonant pumping into the trion state
are shown in Fig.~\ref{fig:exp4}. Their shape does not change
markedly for varying pump densities. Only the long-living electron
spin dephasing is observed, which becomes shorter for higher pump
densities. The corresponding dephasing times are plotted in
Fig.~\ref{fig:exp4t}. Two characteristic regions are seen in
Fig.~\ref{fig:exp4t}: a strong decrease from 14~ns down to 4~ns at
low densities, and a much slower decrease for pump densities
exceeding 100~W$/$cm$^2$. The decrease of the dephasing time can
be understood in terms of delocalization of the resident electrons
caused by their heating due to interaction with photogenerated
carriers. Thus, the electron localization in the studied samples
favors longer spin dephasing times. The change of the behavior of
spin dephasing times at $P\sim 100$~W$/$cm$^2$ can be attributed
to saturation of the trion absorption: a further increase of the
pumping intensity does not change strongly the number of
photogenerated carriers.

\begin{figure}[htb]
\centering
\includegraphics[width=0.95\linewidth]{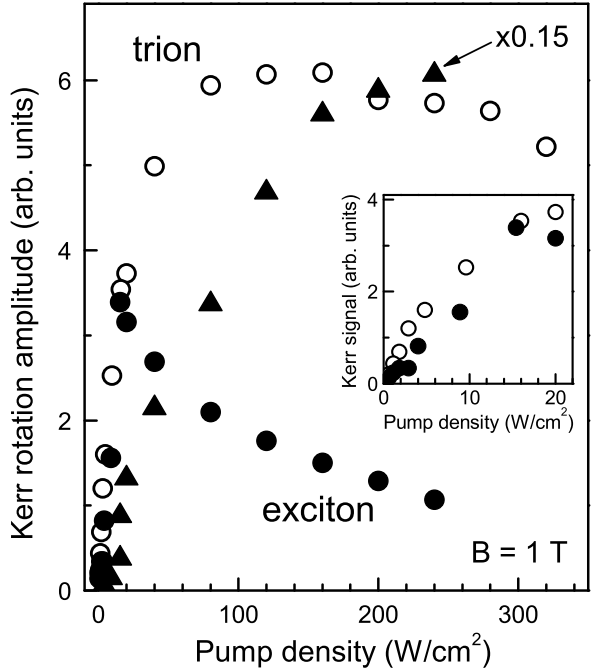}
\caption{Kerr rotation amplitude versus pump density for the
experimental data of Figs.~\ref{fig:exp3} and \ref{fig:exp4}.
Results for the pump resonant with the excitons and trions are
shown, respectively. The amplitude of the fast decaying component
evaluated for zero delay and multiplied by a factor 0.15 is given
by the triangles. The amplitudes of the long-lived 2DEG coherence
measured at a delay of 0.5 ns are shown by open and closed circles
for trions and excitons, respectively. The inset highlights the
low density regime. $T=1.9$~K.}\label{fig:exp5}
\end{figure}

\begin{figure}[htb]
\centering
\includegraphics[width=1\linewidth]{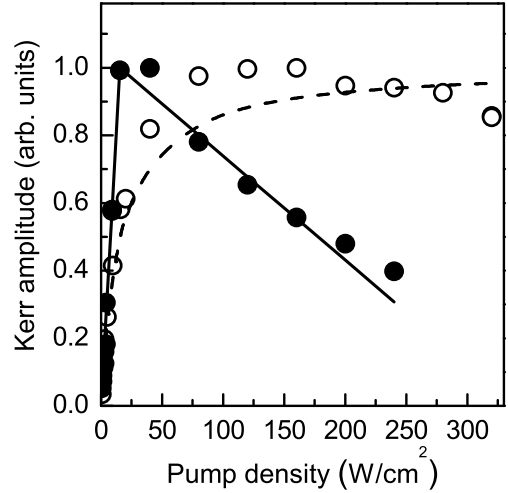}
\caption{Normalized long-lived amplitude of the 2DEG Kerr rotation
shown in Fig. ~\ref{fig:exp5} measured at a delay of 0.5 ns under
resonant pumping of the excitons (closed circles) and the trions
(open circles). Model calculations are shown by the lines (the
dashed curve has been calculated for trion resonant excitation
according to Eq. \eqref{trion_res_sat}. The solid lines have been
calculated for excitation at the exciton frequency according to
Eqs. \eqref{se_low}, \eqref{se_exc}. In the latter case, the
fitting parameter was the ratio of $\tau_s^X/\tau_0$ which is
found to be 10.}\label{fig:exp5m}
\end{figure}

Kerr rotation amplitudes of the 2DEG measured at the trion and
exciton resonances as functions of pump density at $B=1$~T are
shown by the circles in Fig.~\ref{fig:exp5}. The signal has been
measured at a delay of 0.5~ns, when the contribution of the fast
decaying component is vanishingly small. At low excitation density
both dependencies show to a good approximation a linear behavior
as is illustrated by the inset. At higher excitation density the
Kerr rotation amplitudes demonstrate pronounced non-linear
behaviors: for the ``pump/probe T'' configuration saturation is
observed while for the ``pump/probe X'' configuration the KR
amplitudes decreases with increase of the pump power. Both these
results are in agreement with the theoretical predictions of Secs.
\ref{trion} and \ref{exciton}, respectively, as illustrated by the
Figs.~\ref{fig:satur:theor} and \ref{fig:excexc:theor}. The
amplitude of the fast decaying component related to the
electron-in-exciton precession in given in Fig.~\ref{fig:exp5} by
triangles. Note the scaling factor 0.15, which shows that this
component dominates over other signals at moderate and high pump
densities (see also Fig.~\ref{fig:exp3}).

In order to make a quantitative comparison we replot the
experimental points in Fig.~\ref{fig:exp5} using a different
vertical scale, see Fig.~\ref{fig:exp5m}. Namely, we normalize the
Kerr signals to their maximum values (which, according to our
theoretical predictions, correspond to the electron spin density
being equal to $n_e/4$). This allows us to compare the efficiency
of spin coherence generation. One can see that in the low pumping
regime the spin coherence generation efficiency per absorbed
photon is practically the same for the laser tuned either to the
exciton or the trion resonance. This is in a good agreement with
our theory: each absorbed photon creates a trion either directly
or via an intermediate excitonic state, thus an electron with a
given spin orientation is removed from the 2DEG. The strong
pumping regime is different for exciton and trion excitation. In
the case of the laser tuned to the trion resonance the spin of the
2DEG saturates (the small decrease of the Kerr rotation amplitude
can be attributed to heating of the 2DEG) while for the laser
tuned to the exciton resonance a strong decrease of the spin
coherence generation efficiency is seen.

The curves in Fig.~\ref{fig:exp5m} are the results of theoretical
calculations based on the models outlined in Secs. \ref{trion} and
\ref{exciton}. The dashed curve corresponds to trion resonant
excitation, while the solid line is for exciton resonant
excitation. For the dashed curve the only fitting parameter was
the saturation level. For the solid line the only fitting
parameter was the ratio between the electron-in-exciton spin
relaxation time and the exciton radiative lifetime,
$\tau_s^X/\tau_0^X$. The best fit corresponds to a
$\tau_s^X/\tau_0^X=10$.

\subsection{Temperature dependence}

\begin{figure}[htb]
\centering
\includegraphics[width=0.95\linewidth]{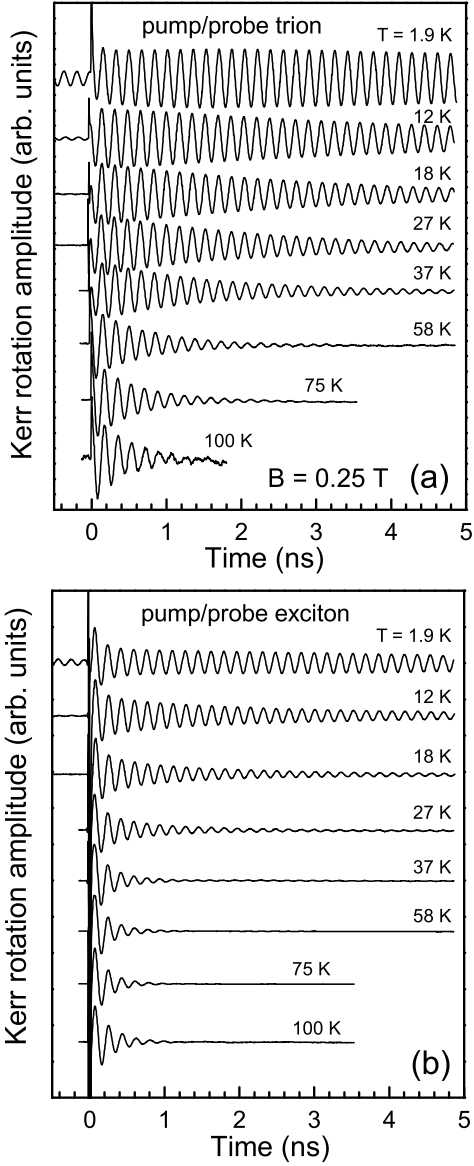}
\caption{Kerr rotation signal measured at different temperatures
by degenerate pump-probe resonant with the trion (a) and the
exciton (b). The signals are normalized on their amplitudes at
zero delay. Pump density $5$~W$/$cm$^2$ and probe density
$1$~W$/$cm$^2$. }\label{fig:exp6}
\end{figure}

\begin{figure}[htb]
\centering
\includegraphics[width=1\linewidth]{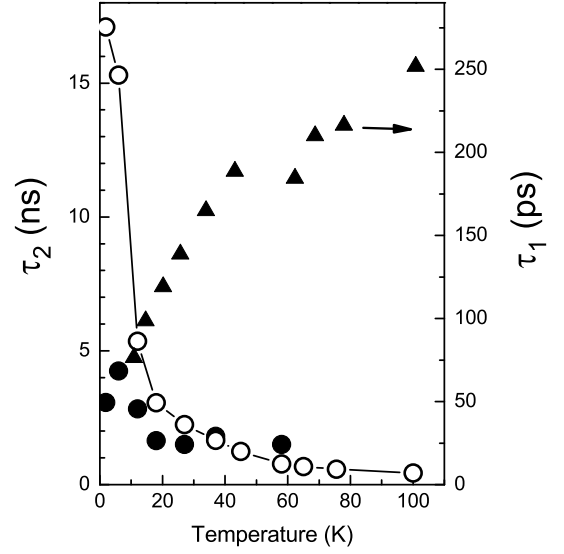}
\caption{Decay times determined from the Kerr rotation signals
measured at different temperatures in Fig.~\ref{fig:exp6}. The
data evaluated from a single exponential fit to the ``pump/probe
trion''\ data are shown by the open circles and a line as a guide
to the eye. The data for the ``pump/probe exciton''\ configuration
are given by the closed symbols: by the circles for $\tau_2$ and
by the triangles for $\tau_1$. $B=0.25$~T, pump density
$5$~W$/$cm$^2$ and probe density $1$~W$/$cm$^2$.
}\label{fig:exp6t}
\end{figure}

The electron spin coherence in CdTe/(Cd,Mg)Te QWs is robust
against temperature increases and can be clearly traced up to
100~K. Kerr rotation signals measured at the trion and exciton
resonances in a temperature range from 1.9 up to 100~K are
presented in Fig.~\ref{fig:exp6}. The signals are normalized to
their values at zero time delay in order to highlight the trends
in signal decay with increasing delay. One can see, that the decay
of the spin beats is thermally accelerated both for resonant
pumping in the trion and in the exciton resonance, but in the
former case the decrease is slower. The decay times are shown in
Fig.~\ref{fig:exp6t}. The Kerr rotation signals for the
``pump/probe trion'' configuration were fitted by a single
exponential decay, and the resulting decay times are given in the
figure by the open circles. A relatively long $T_2^*$ time of
440~ps is measured at $T=100$~K at the trion resonance. The
signals for the ``pump/probe exciton'' configuration were fitted
by double exponential decays, see Eq.(\ref{st100}), for
temperatures below 60~K. Above $T=60$~K only the fast component
with a decay time in the $200-250$~ps range is apparent. This time
can be assigned to exciton recombination and we may conclude that
for high-temperatures the exciton spin coherence dominates over
the 2DEG signal. This can be explained by a reduction of the trion
formation from excitons when the 2DEG electrons have elevated
kinetic energies.

\begin{figure}[htb]
\centering
\includegraphics[width=1.05\linewidth]{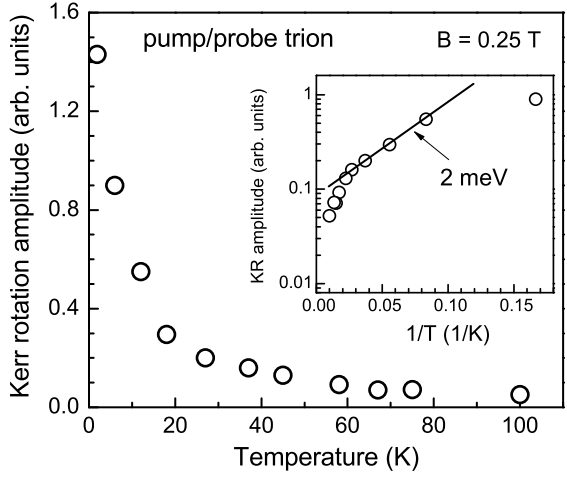}
\caption{Temperature dependence of the Kerr rotation amplitude of
the 2DEG signals in Fig.~\ref{fig:exp6} (a). The inset shows a
logarithmic plot of the amplitude on inverse temperature. The line
corresponds to an activation energy of $2$~meV.}\label{fig:exp7}
\end{figure}

We turn now to the analysis of the Kerr rotation amplitude. Its
temperature dependence for trion resonant pumping is plotted in
Fig.~\ref{fig:exp7}. The amplitude of the first maxima after the
pump pulse, which closely coincides with the amplitude obtained by
extrapolation to zero delay is plotted. In the main panel the Kerr
rotation amplitude is shown as a function of temperature on a
linear scale. The signal rapidly looses about $80\%$ of its
intensity by a temperature increase from 1.9 up to 20 K and then
gradually decreases in intensity up to 100 K, which is the maximum
temperature at which signal could be recorded in experiment. The
insert shows the data in a form which allows extraction of
characteristic activation energies in the temperature dependence.
In the temperature range from 12 to 45 K we obtain an activation
energy of $2$~meV, which equals to the trion binding energy.

The strong temperature dependence of the measured signals can be
explained by two mechanisms: (i) delocalization of electrons and
(ii) dissociation of trions. At very low temperatures $k_B T
\lesssim 0.5-1$~meV, both electrons and photogenerated trions are
localized in the quantum well plane by alloy fluctuations and/or
interface roughnesses. The Kerr rotation signal at fixed pumping
is proportional to the number of photogenerated trions. Increase
of the temperature up to $k_B T = 0.5-1$~meV is accompanied by
electron delocalization which in turn leads to a decrease of the
interaction between photogenerated electron-hole pairs and
electrons. Thus, the number of photogenerated trions decreases.
Further, the increase of the temperature leads to thermal
dissociation of trions, whose number is proportional to
$\exp{(-E_B^T/k_BT)}$, where $E_B^T\approx 2$~meV. Thus, the trion
contribution to the KR signal becomes weaker. At temperatures
strongly exceeding $E_B/k_B$ the main channel for spin coherence
formation is due to exciton-electron spin-flip scattering, see
Sec. II. D.

\subsection{Initial phase shift}

\begin{figure}[htb]
\centering
\includegraphics[width=0.9\linewidth]{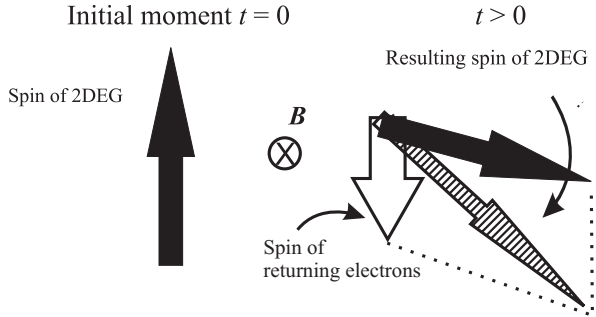}
\caption{Diagram explaining the phase shift of the Kerr rotation
signal.}\label{fig:Aa2}
\end{figure}

The total spin of the electron ensemble precesses around the
external magnetic field. The decay of the total spin with
increasing delay is governed by the spin dephasing processes. The
spins of the electrons forming a trion are in a singlet state and
therefore do not precess around the magnetic field as the resident
electrons do. Recombination of trions leads to return of the
partially $z$-polarized electrons to the 2DEG. Their spin
orientation differs from that of the precessing electrons, which
results in a shift of the initial phase of the measured KR signal,
see Eq. \eqref{sza}. This effect is illustrated in
Fig.~\ref{fig:Aa2}, where the full arrow shows the total spin of
the 2DEG, while the open arrow shows the spin of the returned
electrons. One can see that after trion recombination the total
spin of the electron gas has been rotated by a larger angle as
compared with the rotation of the resident electrons spin. This
induces a phase-shift of the oscillating Kerr signal.

\begin{figure}[htb]
\centering
\includegraphics[width=1\linewidth]{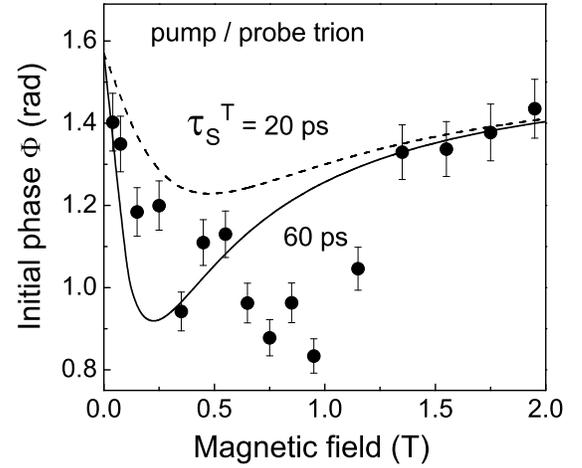}
\caption{Initial phase of the Kerr rotation spin beats versus
magnetic field: (a) experimental data measured for degenerate
pump-probe resonant with the trion. Pump density $0.64$~W$/$cm$^2$
and probe density $0.5$~W$/$cm$^2$. The model calculations have
been performed according to Eq.(\ref{phaseNEW}). }\label{fig:exp8}
\end{figure}

We have analyzed the initial phase by extrapolation of the spin
beats at longer delays back toward zero delay. The experiments
were performed at low excitation density with pump/probe at the
trion resonance. The results are given in Fig.~\ref{fig:exp8}. A
pronounced minimum is seen at $B=0.6-0.8$~T in qualitative
agreement with the model calculations shown by the solid line,
which was calculated with a trion lifetime $\tau_0^T=30$~ps. We
also assumed an electron spin dephasing time $\tau_s=T_2^*=2$~ns
(which corresponds to the experimental value at $B=3$~T). In
experiment this time is longer at smaller fields. However, the
calculated curve is insensitive to the choice of this value as far
as $\tau_s$ remains the longest time in the system. The only free
parameter in the calculations was the spin relaxation time of the
hole in a trion $\tau_s^T$. The two curves correspond to
$\tau_s^T=20$~ps (dashed) and $60$~ps (solid). It is seen, that in
accordance with Eqs. \eqref{phase_min} and \eqref{omega_min}, the
depth of the mininum and its field position is controlled by
$\tau_s^T$. The model results demonstrate qualitative agreement
with the experimental data.

\subsection{Excitons and trions probing different resident electrons}

As we have already noted above, the studied sample contains a
diluted 2DEG with the Fermi energy smaller than the typical
localization potential caused by well width fluctuations. As a
result, at low temperatures of both the lattice and the electron
gas, a major fraction of electrons is localized. We have concluded
also from the pump density and temperature dependencies that the
dephasing time $T_2^*$ is strongly sensitive to electron
localization, see Figs.~\ref{fig:exp4t} and \ref{fig:exp6t}.
$T_2^*$ can be extremely long for the localized electrons and
shortens with electron delocalization.

\begin{figure}[htb]
\centering
\includegraphics[width=1.1\linewidth]{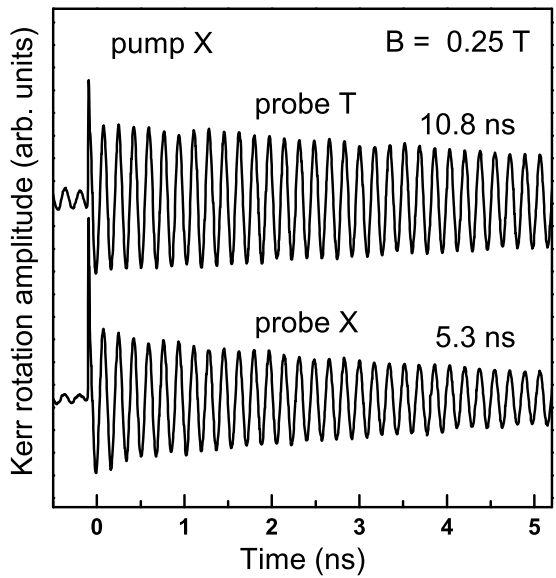}
\caption{Kerr rotation measured by two-color pump-probe.
$T=1.9$~K. Pump density $1$~W$/$cm$^2$ and probe density
$0.4$~W$/$cm$^2$.}\label{fig:exp9}
\end{figure}

To have a deeper insight into the effects of electron localization
on the spin coherence we have done two-color pump-probe
measurements in the regime where the longest dephasing times have
been achieved, i.e. using a low pump density in a weak magnetic
field of 0.25~T. Figure~\ref{fig:exp9} shows the results of such
an experiment, where the pump beam is resonant with the exciton
transition and the Kerr rotation signal is detected at either the
trion or the exciton energy. Under these experimental conditions
the fast decaying component is very small for probing at the
exciton energy. The excitation conditions are identical for both
signals in Fig.~\ref{fig:exp9}, therefore one can expect to detect
the same dephasing times for the 2DEG, irrespective of the probe
energy. Therefore, it is rather surprising to observe that the
dephasing times of the electron coherence differ by a factor of
two: $T_2^*=10.8$ and 5.3~ns for probing at the trion and the
exciton resonance, respectively.

To explain this difference we suggest that different fractions of
the resident electrons contribute to the Kerr rotation signal
measured at the trion or the exciton energies. This can be
understood if we take into account that different mechanisms lead
to Kerr rotation signal at the exciton and trion energy. For
detection at the trion resonance the effect is contributed by the
variation of the trion oscillator strength, which is directly
proportional to the concentration of electrons with a specific
spin orientation, see Sec. II. A. As the in-plain localization is
important to achieve stability of the trions in QWs one may expect
that the trions are much more dependent on the localized
electrons, which have a longer dephasing time. The Kerr rotation
signal at free exciton energy monitors 2DEG spin beats mostly due
to the spin-dependent exciton-electron scattering, see Sec. II. C.
and Ref.~\cite{Ast00}. Possibility for this scattering to occur
implies existence of free electrons. Therefore, at the exciton
energy we address free or quasi-free resident electrons which have
shorter dephasing times.

\section{Conclusions}

We have demonstrated experimentally the possibility to excite spin
coherence of a 2DEG by resonant pumping into the trion or the
exciton states, as well as by non-resonant excitation. It was
shown that the time-resolved Kerr rotation signal detected
experimentally at the trion and exciton frequencies contains two
components: a fast contribution (which vanishes $\sim 30$~ps after
the pump pulse) and a long-living one (with typical decay times of
the order of nanoseconds). The fast component is related with
electrons-in-excitons, while the long one is due to the resident
electrons of the 2DEG. Experimentally we can clearly separate
these contributions.

A theoretical model has been developed to describe various
scenarios of spin coherence generation in QWs with a 2DEG. It is
based on a classical approach to spins and accounts for resonant
excitation in the trion and exciton resonances and also for
nonresonant photoexcitation. A comprehensive set of experimental
results is consistently described in the frame of this model.

The suggested model can be generalized to the description of spin
coherence generation in a 2D hole gas. In this case the in-plane
component of the heavy-hole $g$-factor is negligible and,
therefore, the regime of weak magnetic fields is realized. It can
be expected that due to the stronger spin orbit interaction the
free hole spin relaxation is faster than that of free electrons.
On the other hand, in a low density hole gas most of the carriers
are localized, so that the role of the spin orbit interaction is
diminished and the hole spin relaxation times can reach up to
600~ps in GaAs/(Al,Ga)As QWs \cite{Syp07}, for example.

It is also worthwhile to note here, that the results of our model
consideration can be also applied to for describing the excitation
of spin coherence in other low-dimensional semiconductor
structures such as in an ensemble of singly-charged quantum dots.

\acknowledgements

We acknowledge fruitful discussions with Al. L. Efros, A. Shabaev,
I. V. Ignatiev and I. A. Yugova. This work was supported by the
BMBF "nanoquit" program, the Deutsche Forschungsgemeinschaft
(grant No. YA 65/5-1) and by the Russian Foundation for Basic
Research. EAZh stays in Dortmund were financed by the Deutsche
Forschungsgemeinschaft via grants 436RUS17/79/04, 436RUS17/93/05
and 436RUS17/77/06. An ELI research visit to Dortmund was
supported by the Gambrinus guest-professor program of the
Universit\"at Dortmund. MMG was partially supported by the
``Dynasty'' foundation---ICFPM.

\end{document}